\documentclass[useAMS]{mn2e}
\usepackage{times}
\usepackage{psfig}

\def\ltsima{$\; \buildrel < \over \sim \;$}
\def\lsim{\lower.5ex\hbox{\ltsima}}
\def\gtsima{$\; \buildrel > \over \sim \;$}
\def\gsim{\lower.5ex\hbox{\gtsima}}
\newcommand{\f}{\frac}

\newcommand{\ep}{\epsilon}

\begin{document}

\title[Possible evidence of surface vibration...]{\bf Possible evidence of surface vibration of
strange stars from stellar observations.}

\author[Ray, Dey, Dey \& Bhowmick]
{ Subharthi Ray $^{1,2,3}$   Jishnu Dey $^{1,4}$,
Mira Dey $^{1,5}$ \& Siddhartha Bhowmick, $^{1,6}$ \\
$^1$ Inter University Centre for Astronomy and Astrophysics, Post
Bag 4, Pune 411007, India \\ $^2$ The Abdus Salam International
Center for Theoretical Physics, 34100 Trieste, Italy\\ $^3$
Instituto di Fisica, Universidade Federal Fluminense, Niteroi
24210-340, RJ, Brasil \\ $^4$ UGC Research Professor, Dept. of
Physics, Maulana
Azad College, 8  Rafi Ahmed Kidwai Road, Kolkata 700 013, India\\
$^5$Dept. of Physics, Presidency College, 86/1 College
Street, Kolkata 700 073, India\\
$^6$ Department of Physics, Barasat Govt. College, Barasat, North
24 Paraganas, W. Bengal, India. }

\maketitle

\begin{abstract}
{Emission lines in the $eV$ and $keV$ range by certain stellar
candidates from their recent analysis invoke the question of
their possible origin. These stars under consideration, are the
4U~0614$+$091 (0.65, 0.86, and 1.31 $keV$), 2S~0918$-$549 (0.8
$keV$ with width 55 $eV$), 4U~1543$-$624 (0.7 $keV$), 4U~1850
$-$087 (0.7 $keV$) and 4U~1820$-$30 (0.6 and 0.9 $keV$) and also
the 0.6 $keV$ excess emission in RX~J170930.2$-$263927. Recently,
it has been suggested that the resonance absorption at $\sim$ in
0.7, 1.4, 2.1 and 2.8 $keV$ 1E1207$-$5209 and 0.35, 0.7 and 1.4
$keV$ RX~J1856.5$-$3754 are due to harmonic surface vibrations in
strange stars. We propose that these harmonic vibrations may also
responsible for emission lines in the above mentioned compact
stellar candidates. }

\end{abstract}


\section{Introduction}

Line emissions below and above the keV region have been observed
in many compact star sources. It is intriguing to note that often
their frequencies are multiples of each other - showing a
harmonic origin. It is well known that such harmonic vibrations
are seen in large nuclei and are called breathing mode
oscillations whose origin is the compressional modes of surface
vibrations. To see such behaviour in a compact star, it should
have a sharp surface and the star should be self bound with
pressure p = 0 at the surface. For such criterion, it is required
that the star be described by an equation of state (EOS) which
has a minimum value of E/A at a non-zero density, where E is the
binding energy and A is the baryon number. All of the existing
neutron star EOSs have surface density zero, while for most of the
strange star EOSs have the feature of sharp surface. Besides the
existing MIT Bag model EOSs for strange matter (Farhi \& Jaffe,
1984; Haensel et al., 1986; Alcock et al., 1987, etc.),  Dey et
al. (1998)(henceforth, D98) gave an EOS for strange matter and
used it to model strange stars having such property. Another
strange star model having such phenomena is by Malheiro et al.
(2003), where they have the feature of sharp surface. We shall
use the D98 model in our further discussions and shall consider
only the EOS1 in D98 which is the same as the SS1 in Li et al.
(1999a). Recently, it has been shown that the resonance
absorption in 1E~1207.4$-$5209 and RX~J1856.5$-$3754 can be
plausibly interpreted as being due to surface vibrations in
strange stars (Sinha et al., 2003 - henceforth called as SDDRB).
In particular, the former star shows evidence of 4 harmonics
(Bignami et al. 2003) and has been studied extensively by
XMM-Newton making it the most deeply scrutinized galactic target
of the mission (De Luca et al. 2003).

 The presence of the vibrational frequencies we discuss, is
related to surface compressional modes and it is only possible if
(uds) quark matter is self-bound as for example in D98. There
cannot be harmonic compressional modes in neutron stars because
of the lack of a minimum in the E/A of its EOS which results in
the absence of a sharp surface.

The direct estimate of the mass-radius relation of the compact
stars are still far from accuracy and depends on a lot of {\it
assumptions}, like the surface temperature, luminosity, distance
estimates, etc. With the size of the conventional neutron stars
being slightly more than the more exotic strange stars, and they
themselves being so far away, it is hardly possible to make any
distinction between the two. Ambiguities arise in the estimates
of these objects even with the most modern satellites and
telescopes. A particular example is the recently debated object
RX~J1856.5$-$3754. Drake et al. (2002) claimed it to be a strange
star, while at the same time Walter and Lattimer (2002) showed
that if they assumed the surface temperature of 33 eV, instead of
61.2 eV as taken by Drake et al. (2002), then it gives a
mass-radius relation, perfectly consistent with the normal
neutron stars. So, to bypass these kind of uncertainties and
debates, an alternative signature is desperately sought in order
to prove the existence of the strange stars. The estimates of the
surface vibrations for the strange stars (for the D98 model,
which is a realistic model of uds quark confinement and
asymptotic freedom) has recently been studied by Sinha et al.
(2003). Their predictions of the emissions from the bare strange
stars, due to this effect, astonishingly match with the line
emissions from certain candidates.

In the next  section we  talk about the  radial breathing mode
oscillation problem. In  section 3 we discuss our results and
remark on  relevant observations.  In the last section we conclude
and summarize.

\section{ Breathing oscillations in strange stars. }

The u and d quarks are believed to be light (4 and 7 $MeV$
respectively) and the strange quark s, moderately light (150
$MeV$) at very high density, whereas in hadrons they have a mass
of roughly one third of the hadron  mass. One therefore thinks of
a chiral symmetry restoration as one moves from low to high
density. When this is included along with a Debye screening for
gluon propagation, a model may  be described as a realistic one
for exploring the possibility of a high density strange quark
phase comprising of $ u, ~d ~and ~s$ quarks. Considering the above
mentioned properties, D98 developed their model for strange quark
matter and strange star. The EOS1 of D98 has a minimum of energy
per baryon (E/A) at a surface density of 4.586 times the normal
nuclear matter density, where the pressure $p=0$, and this
surface can vibrate. The spectrum of the vibration frequency $\nu$
is controlled by $dp/dr$ and is harmonic so frequencies $n\nu$
are expected to occur for $n~ = ~ 1,2,3...$.

We recall the relevant formulae for breathing surface
oscillations in SDDRB to calculate the mass, once the fundamental
frequency of vibration is known. We get the corresponding  radius
of the star from the solutions of the Tolman - Oppenheimer -
Volkof (TOV) equation for the star given in D98 and Li et al.
(1999a \& 1999b).

\begin{figure}
\centerline{\psfig{figure=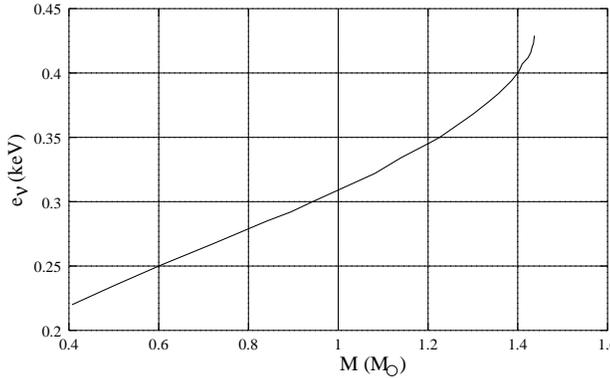,width=8cm}}
\caption{Fundamental vibrational energy (e$_\nu$) for skin vibrations of strange stars
as function of the  stellar mass.}
\label{oscl}
\end{figure}

In D98 model, the energy per particle has a minimum at the
surface. The nature of the curve near the minimum can be
approximated by the potential of a harmonic oscillator. A Taylor
expansion of the energy about $r= R$ gives

\begin{equation}
\f{E(r)}A ~=~  \f{E(R)}A + \f12 k(r)~~ (r-R)^2
\end{equation}
where R is the star radius,
\begin{equation}
k(r)~=~-4\pi~r^2~\f{dp}{dr}
\end{equation} and

\begin{equation}
\f{dp}{dr}~=~ \f{-G(p(r)+\ep(r))(m(r) + 4\pi~r^3
p(r))}{r^2~(1-\f{2Gm(r)}{r})}
\end{equation}
is the TOV equation with conventional notation.

The frequency of the vibration is given by

\begin{equation}
\nu~= ~\f{c}{2 \pi} \sqrt\f{k(R)}{m_{skin}},
\end{equation}
where the mass of the skin ($m_{skin}$) is in energy units.

The $m_{skin}$ is estimated to be very small (SDDRB). The very
small depth of the surface $d$ must be larger than the effective
diameter of the ud quarks $D_{eff}$ :  $d \ge D_{eff}$.

For the extreme case when the equality sign holds, which means the
layer is just one diameter thick, we can replace $D_{eff}$ by $d$.
This yields an estimate of the Maxwellian mean free path :

\begin{equation}
\lambda~ = ~\f{1}{\sqrt{2}~ \pi~ \rho_{surf}~ d^2}~= ~4.28~
\rm{mm},
\end{equation}
where the surface density $\rho_{surf} =  4.586~ \rho_q$ for SS1
with $\rho_{q}~=~ ~ 0.51/fm^3$. The dimension of $ \lambda$ is
clearly macroscopic. It shows the onset of asymptotic freedom
even at the surface of the strange star.

In terms of this  $d$

\begin{equation}
m_{skin} ~= ~ 4 \pi R_{star}^2  ~  d~ \rho_ {surf} \f{M_u +
M_d}{2}.
\end{equation}
${M_u} =  132$ and $M_d = 135$ $MeV$ for SS1 at the star surface.
Mark that this shows chiral symmetry restoration is setting in
already at the surface of the star, being much more prominent at
the centre where the density is 15 $\rho_0$ or 46.85 $\times
10^{14} ~\rm{ gm/cm}^3$.

\begin{figure}
\centerline{\psfig{figure=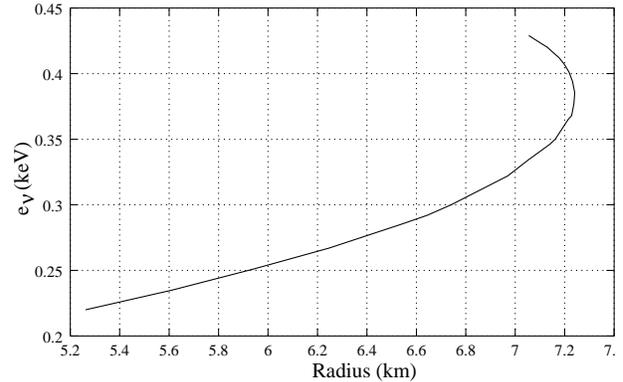,width=8cm}}
\caption{Fundamental vibrational energy (e$_\nu$) for skin
vibrations of strange stars as function of the stellar
dimensions.}
\label{oscl2}
\end{figure}

Due to  strong interaction binding, the strange star surface  is
very firm and the portion  that can oscillate symmetrically is
thin, typically a one quark layer as we have discussed above.

The star mass and radius are given in Figs. (\ref{oscl} \&
\ref{oscl2}) for the relevant value of the breathing resonance
line.

\section{Discussion.}


Below we discuss the evidence for
the line emissions from various stars~:

\begin{enumerate}

\item  4U~0614$+$091 is a moderately bright X-ray source. Three
emission lines at around 0.65, 0.86, and 1.31 $keV$ are proposed by
Schulz (1999). This would correspond to a mass of about $ 1.1~
M_{\odot}$ with radius 7.02 km.

\item  2S~0918$-$549 is a  source fitted to a power law plus black
body spectrum ($kT~= ~3$ $keV$) with a $\chi^2 ~= ~ 1.65$ and shows a
residual at 0.8 $keV$ with width 55 $eV$. Including this line in the fit
reduces the $\chi^2$ to $1.14$ Schulz (1999). We predict a mass of  $ 1.4~
M_{\odot}$ and a radius of 7.22 km.

\item White et al. (1999) found a 0.7 $keV$ line emission for 4U~1543$-$624.

\item Juett, Psaltis and Chakraborty (2001) suggest that the three
stars above and  4U~1850$-$087 with a 0.7 $keV$ line might have
extra neon in the mass donors. However as these authors point
out, a few type I X-ray bursts have been reported from three of
these sources : two bursts from  4U~0614$+$091, one from
2S~0918$-$549 and three from 4U 1850$-$087. This may indicate
that the donors in these systems are not C-O dwarfs (Juett et
al., 2001). For 0.7 $keV$ lines, we predict a mass of 1.23
$M_\odot$ and a radius of 7.16 km.

\item  4U~1820$-$30, located in the Globular Cluster NGC~6625 is a
binary star with period 685 sec. Schulz (1999) finds two peaking
lines at 0.6 and 0.9 $keV$ which enables us to suggest its mass as  $
0.94~ M_{\odot}$. Note that this star was predicted to be a strange
star in Dey et al. (1998) from the mass-radius curve.

\item  Finally the 0.6 $keV$ excess emission in RX~J170930.2$-$263927 is
reported by Jonker et al. (2003). We predict the mass of this
star to be $\sim ~ 0.94 ~  M_{\odot}$.

\end{enumerate}


Millisecond  pulsars  (MSPs) have  long  been  considered  one
of  the possible endpoints  of low  mass X-ray binary  evolution.
The  star is thought to  be spun up to  a millisecond period by
accretion from its low  mass companion.  In  the last  five years
this  theory has  been confirmed with the identification of three
accretion powered MSP-s the first one being SAX~J1808.4$-$3658
which was hailed as the holy grail of  X-ray pulsar  astronomy
by  van der  Klis (2000). It  was suggested in Li et al. (1999a)
that this may be a strange star.

The  two   other  millisecond   X-ray  pulsars  X1751$-$305  and
XTE~J0929$-$314 may also be  strange stars.  If so  then all
these three stars have a low magnetic field and low rate of
accretion due to their age and they are not expected to feel
strong excitations involving the entire accreting  star surface
like other younger  strongly accreting stars. Hence the absence
of line spectra and QPO-s from these stars is expected.

It may  be recalled that electrons  in strange stars are  not bound by
strong interaction  and can  stay outside the  sharp star  surface. A
thin, negatively  charged shell of about  100 fm may  be formed outside
the surface due to these electrons (Alcock et al., 1986; Alcock, 1991).

The stars SAX~J1808.4$-$3658, X1751$-$305 and XTE~J0929$-$314 all show
highly coherent  pulses whence  their rotation frequencies  are easily
determined to be 401, 435 and 185 Hz respectively.

In trying to explain why  coherent radiation are detected in the
above MSPs, one may use the model  of Titarchuk et al. (2002), for
SAX~J1808.4$-$3658. They suggested  that the X-ray emission
originates in  the comptonization  process  in a  relatively
optically thin  hot region. They  estimated the electron  density
near the surface  of the star to  be about  $3.3 \times
10^{19}/{\rm  cm}^3$.  The  star radius with the maximum  mass
given by the TOV equation for  eos1 can be seen to be 7.055 km
(Dey et al., 1998; Li et al., 1999a \& 1999b).{\footnote {In
strange stars the quark matter is  bound tightly by strong
interaction  so that the star is very  stable against possible
fast rotation  (Gondek-Rosi\'nska et al., 2000; Bombaci et al.,
2000) and vibration (Sharma et al., 2002; Sinha et al., 2002a).
However  the electrons are not affected by the strong interaction
and with a electron Fermi momentum $\sim 30$ $MeV$, many of  them
are energetic and move away  from the surface.}} If we assume the
electron density to be uniform over a depth of 100  fm on top of
the star surface and take the value from Titarchuk et al. (2002)
we get the number of electrons to be $N_e~= ~10^{24}$. There
will  a resultant  positive  charge   Z,  on  the star. Assuming
it is charge neutral  $Z~= ~N_e$  and to  the outside electron
this will appear to be concentrated at the centre producing a
electrostatic  potential $N_e  e^2/R~= ~  421.4$ $MeV$.  This
has  to be larger than  the maximum kinetic energy of the
electrons  at the star surface
\begin{equation}
(\sqrt{(k_f(R)\hbar c)^2+m_e^2~c^4}-m_e~c^2),
\end{equation}
by  a factor $f>1$,  in order  that the  electrons at the  top
of  the Fermi surface do not  escape. Using Titarchuk et al.
(2002) we find $f~ =  ~ 13$. A guess value of $f ~= ~10$ was used
in SDDRB. So the electron number expected by us is justified by
the work of Titarchuk et al. (2002), giving more credibility to
our calculations.

\section{Conclusions and summary}

In conclusion we point out that there are bits and pieces of
evidence in support of the existence of strange stars. For
example, the short range pairing of ud quarks by a few $MeV$ can
provide for events parallel to fusion bursts which release about
5 $MeV$ per event. Whereas the latter last typically for only 10
seconds - the pairing can go on for hours (minutes) after a star
has suffered prolonged accretion and all (some of) the pairs are
broken. This can explain superbursts which are recurrent within 5
years as seen in 4U~1636$-$53. The alternate scenario involving
carbon burning may work for 4U~1820$-$30 but cannot be uniformly
applied to all superbursters and long bursters (Sinha et al.,
2002b). The absorption spectra of 1E~1207.4-5209 (De Luca et al.
2004) is another case in evidence which may turn out to be a
strange star showing resonance absorption (Sinha et al. 2003).

In summary, we  point out that lines observed in many stars at more or
less around 1 $keV$ may be due to harmonic surface oscillations,  if
they are strange stars. We also reaffirm that the millisecond X-ray
pulsars SAX~J1808.4$-$3658 may be a strange star.

Confirmation of the  existence of strange stars would lead  to a
rich interplay between X-ray astrophysics and QCD.

\section*{Acknowledgments}
SR thanks the FAPERJ for research support while being in the
IF-UFF and the ICTP for short visit. JD, MD and SB likes to thank
the financial support of the DST grant no. SP/S2/K-03/2001, Govt.
of India. (Project name  : Changing Interface  of Nuclear,
Particle and Astrophysics) MD and  JD also thank IUCAA for a
short pleasant stay and attending the Workshop on the Provocative
Universe, June 30 - July 2, 2003.




\end{document}